%% file: main.tex
\documentclass{article}

\input{packages.tex}
\input{commands.tex}

\title{Enhancing Data-Assimilation in CFD using Graph Neural Networks}

\author{Michele Quattromini \\ Department of Mechanics, Mathematics and \\ Management, Polytechnical University of Bari,\\ Italy, BA 70126 \\
  LISN-CNRS, Université Paris-Saclay \\
  Orsay, France, 91440 \\ \texttt{michele.quattromini@poliba.it} 
\\ \And
  Michele Alessandro Bucci \\
  Safran Tech, Digital Sciences \& Technologies\\ Magny-Les-Hameaux, France, 78114\\\texttt{michele-alessandro.bucci@safrangroup.com} 
\\ \And
  Stefania Cherubini \\
  Department of Mechanics, Mathematics and\\ Management,
  Polytechnical University of Bari,\\ Italy, BA 70126 \\ \texttt{stefania.cherubini@poliba.it} \\ 
\\ \And
  Onofrio Semeraro \\
  LISN-CNRS, Université Paris-Saclay \\
  Orsay, France, 91440 \\ \texttt{onofrio.semeraro@universite-paris-saclay.fr} }

\begin{document}

\maketitle

\begin{abstract}
We present a novel machine learning approach for data assimilation applied in fluid mechanics, based on adjoint-optimization augmented by Graph Neural Networks (GNNs) models. We consider as baseline the Reynolds-Averaged Navier-Stokes (RANS) equations, where the unknown is the meanflow and a closure model based on the Reynolds-stress tensor is required for correctly computing the solution. An end-to-end process is cast; first, we train a GNN model for the closure term. Second, the GNN model is introduced in the training process of data assimilation, where the RANS equations act as a physics constraint for a consistent prediction. We obtain our results using direct numerical simulations based on a Finite Element Method (FEM) solver; a two-fold interface between the GNN model and the solver allows the GNN's predictions to be incorporated into post-processing steps of the FEM analysis. The proposed scheme provides an excellent reconstruction of the meanflow without any features selection; preliminary results show promising generalization properties over unseen flow configurations.
\end{abstract}

%======================================================

\section{Introduction} Computational Fluid Dynamics (CFD) is a cornerstone for the simulation and analysis of fluid flows across a multitude of scientific and engineering fields. However, the computational complexity associated with solving the governing equations is a persistent challenge, as it requires high-performance computing resources and elaborate numerical techniques. The combination of Machine Learning (ML) techniques and CFD numerical solvers is emerging as a possible solution to overcome these limits (see reviews by \cite{duraisamy2019turbulence} and \cite{brunton2020machine}).\\
In this paper, we consider a data-assimilation problem augmented by a Graph Neural Network (GNN) model. In particular, we aim at computing the meanflow based on known measurements. Standard approaches rely on data-assimilation via adjoint-based optimization \citep{foures}.
Here, we introduce an end-to-end assimilation process, that includes a GNN model capable of predicting the closure term for the RANS equations, namely the Reynolds stress, following the works by \cite{DSS} and \cite{end-to-end_xiao}. The choice of GNN models is justified by the strong interest for this architecture in the fluid mechanics community \citep{lino2023current}.
Indeed, GNN allows to address some of the limitations often found when integrating Deep Neural Networks (DNN) and numerical solvers for physics. First, data hungriness is often a limiting factor when training a DNN, as from an industrial point of view the availability of data can be limited. GNN are known to be more data-parsimonious \citep{hamilton2020graph}. 
\begin{wrapfigure}{r}{0.3\textwidth}
\centering
\includegraphics[width=\linewidth]{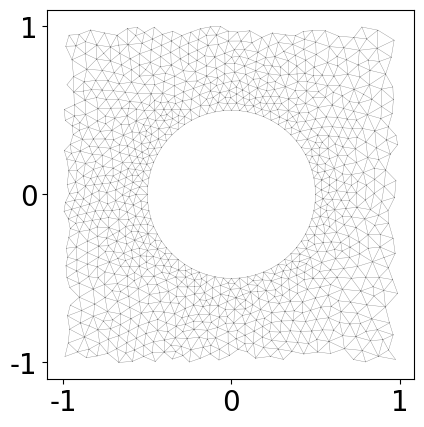}
\caption{Unstructured mesh around a 2D cylindrical body.}\label{fig:1}
\end{wrapfigure}
Second, some DNN architectures combined with numerical solvers lack of flexibility when considering CFD simulations in complex geometries. An example is given by standard Convolutional Neural Networks (CNN), often used as approximators of closure terms, known to face challenges with unstructured meshes. Although very recent works address the latter limitation in CNN \citep{Coscia_2023}, successful works are already available relying on GNN, for instance, for the up-scaling of low-resolution simulations \citep{upscalingCFD-GNN}.

In the following, we briefly introduce our data-assimilation scheme and consider the dynamics of the wake past a cylinder at low Reynolds number, $Re=120$, as a test case. The flow exhibits the von Karman Street instability and is simulated by means of Direct Numerical Simulations (DNS) using a Finite Element Method (FEM) solver. The GNN model is pre-trained on a dataset composed by a single pair of meanflow (input) and Reynolds stress (output) of the wake past cylinder bluff body on unstructured mesh at $Re = 120$. Then, the training iteration proceeds by enforcing the RANS equations in the post-training loop. We show that such a model improves the learning path by effectively assimilating data under the constraint provided by the RANS equations.

%======================================================
%
\section{Methodology}
\subsection{Baseline equations and numerical simulations}
We consider incompressible, two-dimensional (2D) fluid flows developing past bluff bodies. We focus on time-averaged quantities and on second order-statistics, the Reynolds stress. To this end, we introduce the Reynolds decomposition
\begin{equation}
\vdir(\mathbf{x},t) = \mean\vdir(\mathbf{x}) + \fluct\vdir(\mathbf{x},t),\label{eq:reydeco} 
\end{equation}
where $\mean{\vdir}=(\mean{u},\mean{v})^T$ is the meanflow, $\fluct\vdir$ the fluctuation field, and $\mathbf{x} = (x,y)^T$ is the vector composed by the streamwise direction $x$ and the wall-normal direction $y$. Formally, any unsteady flow can be described using this decomposition, whether we consider a laminar case or turbulent one \citep{foures}. Plugging Eq.~(\ref{eq:reydeco}) in the Navier--Stokes equations, after time-averaging we get
\begin{subequations}
    \begin{align}
        \mean\vdir\cdot\nabla\mean\vdir + \nabla\mean\pdir - \invRe\nabla^2\mean\vdir &= \ff\\
        \nabla\cdot\mean\vdir &= 0,
    \end{align}
    \label{eq:RANS}
\end{subequations}
where $\mean{p}$ is the average pressure field. The Reynolds number is defined as $Re={U_{\infty} D}/{\nu}$, with the reference velocity chosen as the free stream velocity $U_{\infty}$ at the inlet, $D$ the reference length of the flow and $\nu$ the kinematic viscosity. Mathematically, the resulting system provides the Reynolds-averaged Navier--Stokes Equations (RANS) and the forcing $\ff$ is the closure term or Reynolds stress. This term can be modeled or -- when data are available -- directly computed as
\begin{equation}
\ff= -\overline{\fluct\vdir \cdot \nabla\fluct\vdir}.\label{eq:forcing}
\end{equation}
In this article, the numerical solution of the Eq.~\ref{eq:RANS}  is performed using a FEM solver based on \verb|FEniCS| \citep{alnaes2015fenics}. Temporal discretization is implemented using a second-order Backward Differentiation Formula (BDF).  FEM solutions are computed on unstructured mesh, thus allowing for a great versatility in handling complex geometries (see Fig.~\ref{fig:1}). The computational domain is discretized using 13283 nodes and has dimensions $[L_x, L_y] = [27, 10]$.

\subsection{Data assimilation problem}
A common problem often found in fluid mechanics is to reconstruct a flow field or a meanflow starting from local measurements. The problem can be cast as an optimization process of data-assimilation, where the Eq.~\ref{eq:RANS} can be used as baseline and the forcing term $\ff$ is the control parameter to be determined. The cost function to be minimized can be written as
\begin{equation}
J(\hat{\mean\vdir}) = \dfrac{1}{2}||(\mean\vdir - \hat{\mean\vdir})||^2,
\label{eq:cost_function}
\end{equation}
where ${\mean\vdir}$ is a known field and $\hat{\mean\vdir}$ is the solution of the RANS equations based on $\hat{\ff}$. The variable $\hat{\ff}$ is the iterative solution of an adjoint-based optimization, based on the procedure introduced in the paper by \cite{foures} and approximating the term $\ff$. The solution is iteratively updated until convergence using the gradient $\nicefrac{\partial J}{\partial\hat{\ff}}$, which corresponds to the adjoint state. Here, this process is augmented through an additional GNN model that provides the closure term.
 
\subsection{Augmented training process} \label{chap:augmented_training}
\begin{figure}[h!]
    \centering
    \includegraphics[width=\textwidth]{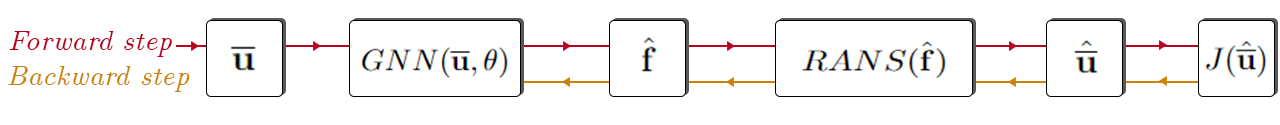}
    \caption{End-to-end training loop; $\mean\vdir$ is the GNN's input meanflow; $\boldsymbol{\theta}$ are the GNN's trainable parameter; $\ff$ is the GNN's predicted forcing; $J(\hat{\mean\vdir)}$ is the cost function as expressed in Eq.~\ref{eq:cost_function}.}
    \label{fig:training_loop}
        \begin{picture}(0,0)
        \put(180,43){$\nicefrac{\partial J}{\partial\hat{\mean\vdir}}$}
        \put(22,43){$\nicefrac{\partial\hat{\mean\vdir}}{\partial\hat{\ff}}$}
        \put(-32,43){$\nicefrac{\partial\hat{\ff}}{\partial\boldsymbol{\theta}}$}
    \end{picture}    
\end{figure}

The training process is inspired by the Deep Statistical Solver framework introduced by \cite{DSS}, and the data assimilation scheme by \cite{end-to-end_xiao}, both developed as end-to-end learning cycles. Our algorithm is sketched in Fig.~\ref{fig:training_loop}, where the \emph{forward step} path and the \emph{backward step} are represented; $\boldsymbol{\theta}$ are the GNN's trainable parameters. Following the Stochastic Gradient Descent (SGD) method and employing the gradients chain-rule, the gradient of the cost function $J$ with respect to $\boldsymbol{\theta}$ is
\begin{equation}
\frac{\partial J}{\partial\boldsymbol{\theta}} = \frac{\partial J}{\partial\hat{\mean\vdir}} \cdot \frac{\partial\hat{\mean\vdir}}{\partial\hat{\ff}} \cdot \frac{\partial\hat{\ff}}{\partial\boldsymbol{\theta}} = \frac{\partial J}{\partial\hat{\ff}} \cdot \frac{\partial\hat{\ff}}{\partial\boldsymbol{\theta}}.\label{eq:backward_step}
\end{equation}
The novelty of this paper lies in the combination of these gradients from diverse computational strategies, achieved through the use of our GNN-FEM interface. In particular, the term $\nicefrac{\partial J}{\partial\hat{\ff}}$ comes from the adjoint formulation of the RANS equations and is computed using the FEM simulation; $\nicefrac{\partial\hat{\ff}}{\partial\boldsymbol{\theta}}$ is obtained from the Automatic Differentiation (AD) package provided by the \verb|Torch| python library. To compute the chain-rule as stated in Eq.~\ref{eq:backward_step}, these terms from different approaches need to be accumulated and back-propagated all together through the entire GNN model. This operation is made possible by the two-folded interface that can convert a FEM discretized vectorial field from \verb|FEniCS| in a numerical tensor to be used by \verb|Torch| and vice-versa, without any loss of information. In the following we will denote the data assimilation scheme as DA-GNN.
%

%======================================================
%
\subsection{GNN architecture description}
A custom GNN architecture has been designed to operate on the unstructured meshes. In the initial phase of the \emph{forward step}, each node within the mesh is assigned with an embedded state, denoted as $\textbf{h}_i$ for each node $i$. The state is initialized to a zero vector. The architecture employs two dedicated Multi-Layer Perceptrons (MLPs) to perform the message-passing mechanism \citep{hamilton2020graph}, one for the in-going messages to the node and one for the out-going information from the node itself. These MLPs extract and integrate the spatial flow information from the neighboring nodes and, together with the embedded state of the node itself of the current iteration, they are processed via a third MLP to provide an update of the embedded state at each node.

The operation reads as
\begin{equation}
\textbf{h}^{(k)}_i = \textbf{h}^{(k-1)}_i + \alpha \Psi^{(k)} \left( \textbf{h}^{(k-1)}_i, \{\mean\vdir, Re\}, \boldsymbol{\phi}^{(k)}_{\rightarrow, i}, \boldsymbol{\phi}^{(k)}_{\leftarrow, i}, \boldsymbol{\phi}^{(k)}_{\circlearrowright, i}\right),
\end{equation}
where ${\boldsymbol{\phi}^{(k)}_{\rightarrow, i} = \frac{1}{N_d} \sum_{j=1}^{N_d} \boldsymbol{\phi}^{(k)}_{j, i}}$ is the message that the node $i$ sends to its {$N_d$} neighbours, averaged over the number of neighbours; ${\boldsymbol{\phi}^{(k)}_{\leftarrow, i} = \frac{1}{N_d} \sum_{j=1}^{N_d} \boldsymbol{\phi}^{(k)}_{i,j}}$ is the message that the node $i$ receives from its {$N_d$} neighbours, averaged over the number of neighbours; ${\boldsymbol{\phi}^{(k)}_{\circlearrowright, i} = \boldsymbol{\phi}^{(k)}_{i, i}}$ the message that the node $i$ sends to itself to avoid  loss of information as the process advances. $\Psi^{(k)}$ is a generic differentiable operator that can be approximated also in this case by MLP; $\alpha$ is a relaxation coefficient that allows to scale each update of the embedded states with the previous one during the message passing loop. At the end of the process, the last embedded state is projected back to a physical state using a decoder, a fully connected MLP, to obtain the closure term prediction of the GNN. We adopted an automatic hyper-parameters landscape explorer, the python library \verb|Optuna|, for tuning the 5 hyper-parameters of the GNN model: the dimension of the embedded state on each node; the number $k$ of GNN layers; the relaxation coefficient $\alpha$ used for the updates; the learning rate and $\gamma$, a coefficient used to weight the contribution to the cost function by each layer of the GNN. 
%======================================================
%
\section{Results}
\begin{figure}[h!]
    \centering
    \includegraphics[width=0.7 \textwidth]{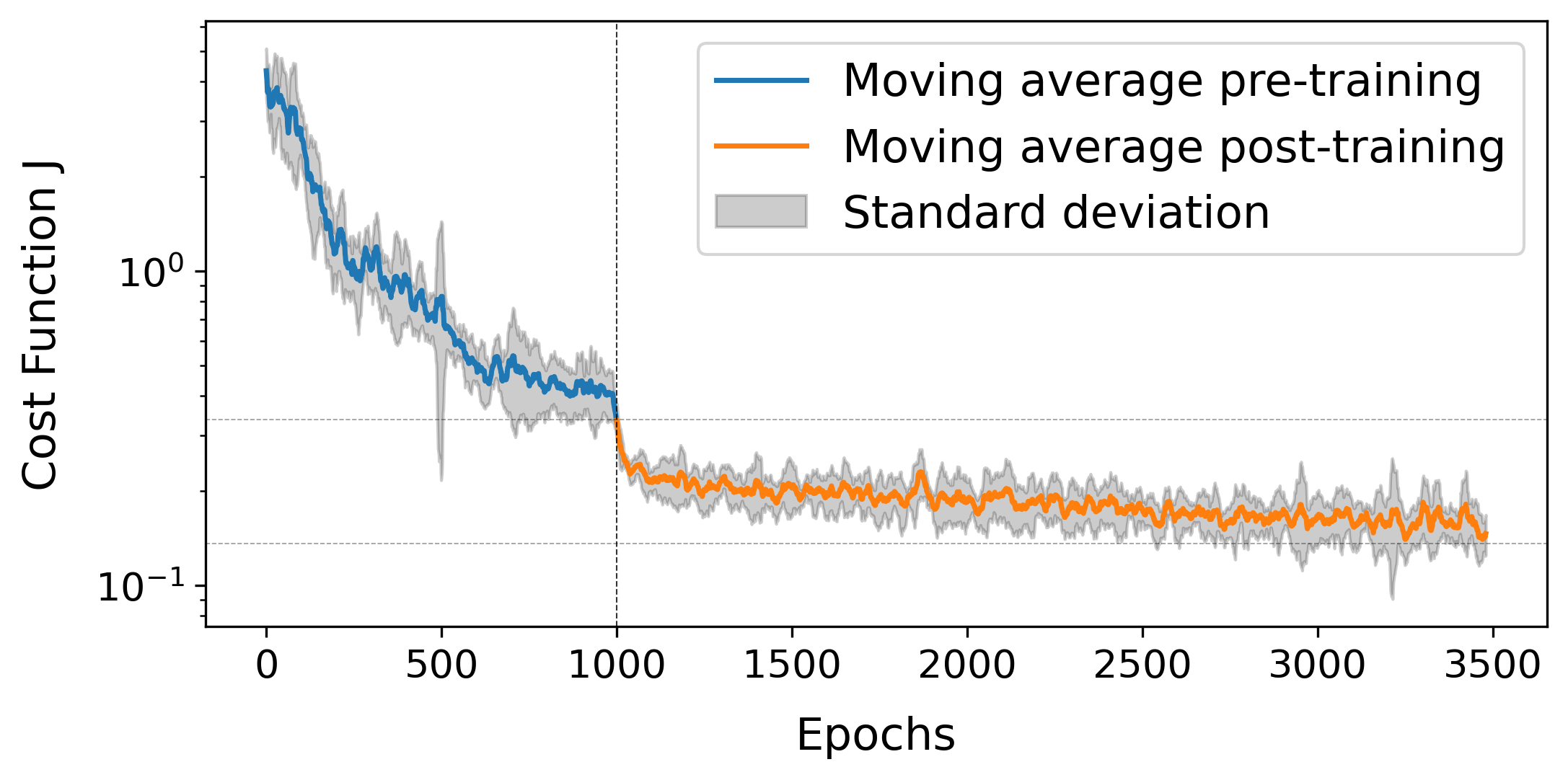}
    \caption{Training loss curve based on the cost function in Eq.~\ref{eq:cost_function}. The blue line (pre-training) and the orange line (post-training) are the moving average of the mean curve from 5 different GNN models, with different random seeds. The grey area shows the standard deviation around the mean curve.}
    \label{fig:loss_curve}
\end{figure}

\begin{figure}[h!]
	\centering
	\includegraphics[width=1 \textwidth]{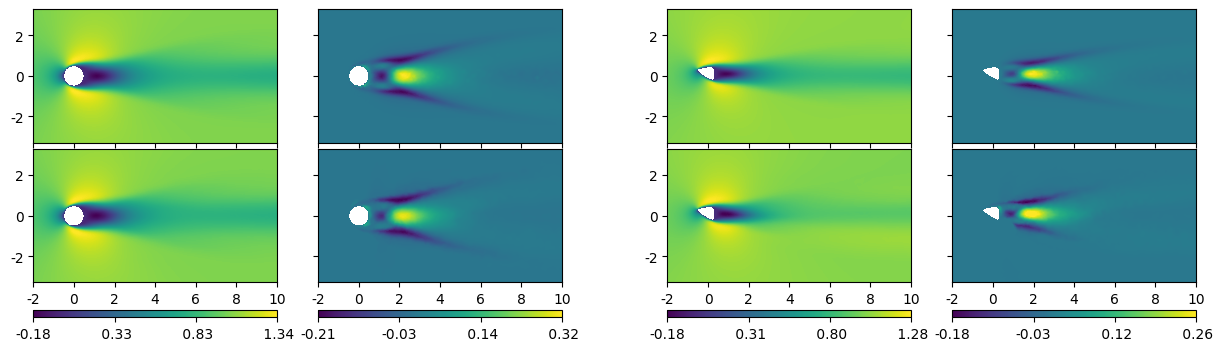}
	\caption{Comparison of the streamwise component between DNS results and DA-GNN predictions; $(a,e)$ $\mean\vdir$ from DNS; $(c,g)$ $\hat{\mean\vdir}$ from DA-GNN; $(b,f)$ $\ff$ from DNS; $(d,h)$ $\hat{\ff}$ from DA-GNN;}
	\label{fig:results}
	\begin{picture}(0,0)
		\put(-220,162){$\textcolor{white}{(a)}$}
		\put(-110,162){$\textcolor{white}{(b)}$} 
  		\put(-220,108){$\textcolor{white}{(c)}$}
            \put(-110,108){$\textcolor{white}{(d)}$}
        
		\put(25,162){$\textcolor{white}{(e)}$} 
		\put(133,162){$\textcolor{white}{(f)}$}
  		\put(25,108){$\textcolor{white}{(g)}$} 
		\put(133,108){$\textcolor{white}{(h)}$}
  
		\put(-240,145){$y$} 
		\put(-240,92){$y$}

		\put(-175,37){$x$}
		\put(-67,37){$x$} 
		\put(67,37){$x$}
  		\put(175,37){$x$}

    \put(-162,180){Training: Cylinder case}
    \put(55,180){Generalization: Random shape case}

	\end{picture}
\end{figure}

We consider as a test case the reconstruction of the meanflow of a cylinder wake at $Re=120$. 
In Fig.~\ref{fig:loss_curve}, the training loss curve is shown. The training process consists of two phases; the first 1,000 epochs are based on a supervised learning of the forcing terms. This initial phase is necessary to provide a starting forcing term for the subsequent phase that allows the RANS solver to converge, addressing FEM numerical stability problems. Therefore, the cost function is designed to minimize the difference between the predicted forcing terms and the actual ones trained over a dictionary of synthetic data obtained for different bluff body shapes \citep{quattromini2023operator}. Then, we move to the full end-to-end training process for another 2,500 epochs, where the cost function $J$ in Eq.~\ref{eq:cost_function} evolves, comparing the predicted meanflow $\hat{\mean\vdir}$ with ground truth from DNS.
The global training loss shows that the cost function is consistently decreasing, assuring that the gradients are correctly computed and back-propagated as the model's predictions are continuously refined. The RANS equations are thus correctly enforced in the training process and the prediction of the DA-GNN holds now physical consistency with the CFD equations. Fig.~\ref{fig:results}($a$-$d$) shows the predicted $\hat{\mean\vdir}$ and $\hat{\ff}$ as compared to the ground truth for the presented cylinder test case, while Fig.~\ref{fig:results}($e$-$h$) shows the generalization capabilities on a random shape bluff body at $Re = 130$, unseen during the training process.\\

%======================================================

\section{Conclusions} In this paper, we introduced a novel algorithm of data assimilation combining the versatility of GNN with a FEM solver. An interface between the GNN model and the FEM solver enables the communication between the two computational environments for a robust back-propagation of gradients during the training phase. The resulting data-assimilation scheme provides an appropriate reconstruction of the meanflow without any features selection, relying on a pre-trained GNN model and keeping physical coherence of the prediction through the constraint provided by the RANS equations. We are currently moving to corrupted or incomplete meanflow reconstruction and refining the technique for enhancing the generalization properties over unseen cases, including different geometries and flow parameters.

%======================================================

\medskip \noindent{\textbf{Acknowledgements.}} The Ph.D. fellowship of M.~Quattromini is supported by the Italian Ministry of University. A part of the research was funded by the grant PRIN2017-LUBRI-SMOOTH of the Italian Ministry of Research and ANR-21-REASON from the French Agency for National Research.

\bibliographystyle{unsrtnat}
\bibliography{biblio}

\end{document}

%% file: commands.tex
\newcommand{\vdir}{\mathbf{u}}
\newcommand{\pdir}{p}
\newcommand{\ff}{\mathbf{f }}
\newcommand{\mean}[1]{\overline{#1}}
\newcommand{\fluct}[1]{#1'}
\newcommand{\invRe}{\dfrac{1}{Re}}

%% file: main.bbl
\begin{thebibliography}{11}
\providecommand{\natexlab}[1]{#1}
\providecommand{\url}[1]{\texttt{#1}}
\expandafter\ifx\csname urlstyle\endcsname\relax
  \providecommand{\doi}[1]{doi: #1}\else
  \providecommand{\doi}{doi: \begingroup \urlstyle{rm}\Url}\fi

\bibitem[Duraisamy et~al.(2019)Duraisamy, Iaccarino, and Xiao]{duraisamy2019turbulence}
K.~Duraisamy, G.~Iaccarino, and H.~Xiao.
\newblock Turbulence modeling in the age of data.
\newblock \emph{Annu. Rev. Fluid Mech.}, 51:\penalty0 357--377, 2019.

\bibitem[Brunton et~al.(2020)Brunton, Noack, and Koumoutsakos]{brunton2020machine}
S.~L. Brunton, B.~R. Noack, and P.~Koumoutsakos.
\newblock Machine learning for fluid mechanics.
\newblock \emph{Annu. Rev. Fluid Mech.}, 52:\penalty0 477--508, 2020.

\bibitem[Foures et~al.(2014)Foures, Dovetta, Sipp, and Schmid]{foures}
D.~P.~G. Foures, N.~Dovetta, D.~Sipp, and P.~J. Schmid.
\newblock A data-assimilation method for {R}eynolds-averaged {N}avier--{S}tokes-driven mean flow reconstruction.
\newblock \emph{J. Fluid Mech.}, 759:\penalty0 404--431, 2014.

\bibitem[Donon et~al.(2020)Donon, Liu, Liu, Guyon, Marot, and Schoenauer]{DSS}
B.~Donon, Z.~Liu, W.~Liu, I.~Guyon, A.~Marot, and M.~Schoenauer.
\newblock {Deep Statistical Solvers}.
\newblock In \emph{{NeurIPS}}, Vancouver, Canada, 2020.

\bibitem[Str{\"o}fer and Xiao(2021)]{end-to-end_xiao}
C.~A. Str{\"o}fer and H.~Xiao.
\newblock {End-to-end differentiable learning of turbulence models from indirect observations}.
\newblock \emph{Theor. App. Mech.}, 11\penalty0 (4):\penalty0 100280, 2021.

\bibitem[Lino et~al.(2023)Lino, Fotiadis, Bharath, and Cantwell]{lino2023current}
M.~Lino, S.~Fotiadis, A.A. Bharath, and C.D. Cantwell.
\newblock Current and emerging deep-learning methods for the simulation of fluid dynamics.
\newblock \emph{Proc. R. Soc. A}, 479\penalty0 (2275):\penalty0 20230058, 2023.

\bibitem[Hamilton(2020)]{hamilton2020graph}
W.~L. Hamilton.
\newblock \emph{Graph representation learning}.
\newblock Morgan \& Claypool Publishers, 2020.

\bibitem[Coscia et~al.(2023)Coscia, Meneghetti, Demo, Stabile, and Rozza]{Coscia_2023}
D.~Coscia, L.~Meneghetti, N.~Demo, G.~Stabile, and G.~Rozza.
\newblock A continuous convolutional trainable filter for modelling unstructured data.
\newblock \emph{Comput. Mech.}, 72\penalty0 (2):\penalty0 253--265, 2023.

\bibitem[Belbute-Peres et~al.(2020)Belbute-Peres, Economon, and Kolter]{upscalingCFD-GNN}
F.~A. Belbute-Peres, T.~Economon, and Z.~Kolter.
\newblock Combining differentiable {PDE} solvers and graph neural networks for fluid flow prediction.
\newblock In \emph{ICML}, pages 2402--2411. PMLR, 2020.

\bibitem[Aln{\ae}s et~al.(2015)Aln{\ae}s, Blechta, Hake, Johansson, Kehlet, Logg, Richardson, Ring, Rognes, and Wells]{alnaes2015fenics}
M.~Aln{\ae}s, J.~Blechta, J.~Hake, A.~Johansson, B.~Kehlet, A.~Logg, C.~Richardson, J.~Ring, M.~E Rognes, and G.N. Wells.
\newblock The {FE}ni{CS} project version 1.5.
\newblock \emph{Archive of numerical software}, 3\penalty0 (100), 2015.

\bibitem[Quattromini et~al.(2023)Quattromini, Bucci, Cherubini, and Semeraro]{quattromini2023operator}
M.~Quattromini, M.~A. Bucci, S.~Cherubini, and O.~Semeraro.
\newblock Operator learning of rans equations: a graph neural network closure model.
\newblock \emph{arXiv:2303.03806}, 2023.

\end{thebibliography}
